\documentclass[twocolumn,aps,amsmath,amssymb,showpacs,superscriptaddress]{revtex4-2} 
\usepackage{epsf}
\usepackage[utf8]{inputenc}
\usepackage{amsmath}
\usepackage{amsfonts}
\usepackage{amssymb}
\usepackage{makeidx}
\usepackage{color}
\usepackage{gensymb}
\usepackage{graphicx}
\usepackage[hidelinks,colorlinks=true,linkcolor=blue,citecolor=blue]{hyperref}

\begin{document}

\title{Complex magnetic interactions and critical behavior analysis in quaternary CoFeV$_{0.8}$Mn$_{0.2}$Si Heusler alloy} 
\author{Guru Dutt Gupt}
\affiliation{Department of Physics, Indian Institute of Technology Delhi, Hauz Khas, New Delhi-110016, India}
\author{P. D. Babu}
\affiliation{UGC-DAE Consortium for Scientific Research, Mumbai Centre, BARC Campus, Mumbai-400085, India}
\author{R. S. Dhaka}
\email{rsdhaka@physics.iitd.ac.in}
\affiliation{Department of Physics, Indian Institute of Technology Delhi, Hauz Khas, New Delhi-110016, India}
\date{\today}
	
\begin{abstract}	
	
We investigate the magnetic behavior and critical exponents of quaternary CoFeV$_{0.8}$Mn$_{0.2}$Si Heusler alloy to understand the interactions across the Curie temperature ($T_{\rm C}$). The Rietveld refinement of the x-ray diffraction pattern with the space group F$\bar{4}$3m confirms a single-phase cubic Y-type crystal structure. The magnetic susceptibility $\chi (T)$ data show a ferromagnetic nature with a second-order phase transition from paramagnetic to ferromagnetic at $446\pm1$~K. The saturation magnetization at 5~K is determined to be 2.2~$\mu_B$/f.u., which found to be close to the Slater--Pauling rule and indicates its half-metallic nature. The values of asymptotic critical exponents ($\beta$, $\gamma$, and $\delta$) and the $T_{\rm C}$ are extracted through detailed analytical analysis including the Modified Arrott plot, the Kouvel-Fisher (K--F) method, and the Widom scaling relation. Interestingly, the estimated values of $\beta$ = 0.369 and $\gamma$ = 1.445 closely approximate the theoretical values of the 3D Heisenberg model and second-order thermodynamic phase transition across the $T_{\rm C}$. The obtained exponents lead to the collapse of renormalized isotherms, represented by the relationship between the magnetization (m) and the applied magnetic field (h), into two distinct branches above and below the $T_{\rm C}$, which validates the reliability of the analysis. Furthermore, these exponents suggest that the spin interaction follows a decay pattern of $J(r) \sim r^{-4.99}$, indicating a short-range magnetic ordering, akin to the itinerant-electron 3D Heisenberg model.

\end{abstract}
\maketitle
	
\section{\noindent ~Introduction}
	
In 2008, X. L. Wang~\cite{Wang_prl_08} suggested a new class of materials reported as spin-gapless semiconductors (SGS) having no band gap in the band structure of one spin channel, but at the same time a finite band gap in another spin channel similar to a semiconductor, at the Fermi level (E$_{\rm F}$) \cite{Wang_prl_08}. This has attracted huge attention of scientific community as it is predicted to show an unusual electronic band structure and serves as an intermediary state between the half-metal and gapless semiconductor states in large number of materials and in particular Heusler alloys \cite{Felser_sips_16, Ouardi_prl_13, Gao_prm_19, Gupt_JVSTA23}. For example, the first experimental evidence of SGS was confirmed in Mn$_2$CoAl \cite{Ouardi_prl_13}, and then many more are predicted to show interesting properties such as Fe$_2$CoSi \cite{Du_epl_13}, Cr$_2$CoAl \cite{Srivastava_aip_20, Gupt_JAP23}, including quaternary Heusler alloys CoFeMnSi \cite{Yamada_prm_18, Alijani_prb_11, Yamada_prb_19, Bainsla_prb_15}. More interestingly, many of these Heusler alloys have attracted a lot of attention in recent years for a wide range of applications in spintronics and heat engines because of their unique properties such as half metallicity, spin gapless, compensated ferrimagnetism, and magneto-caloric properties \cite{Nehla_prb_19, Nehla_jap_19, Srivastava_aip_20}. In this direction, the quaternary Heusler alloys (XX$^\prime$YZ), where X, X$^\prime$ and Y are transition metals, and Z is group element, can be synthesized in LiMgPdSn-type structure or also known as Y-type structure associated with the space group F$\bar{4}$3m \cite{Felser_sips_16, Ren_jsnm_16, Alijani_prb_11, Ozdogan_jap_13}. In this crystal structure the elements in four interpenetrating {\it f.c.c.} sub-lattices occupy all the atomic positions and the preferred occupation of the sites depend on the respective electro-negativity and atomic sizes of the constituent elements \cite{Yamada_prm_18, Kharel_aip_22}. The calculated site preference energies for the quaternary Heusler alloys found three non-degenerate possible atomic configurations (type--I, type--II, and type--III), which reveals the phase stability \cite{Bainsla_prb_15, Rani_prb_17, Alijani_prb_11, Yamada_prm_18, Mishra_jmmm_22, Gao_prm_19} and energetically most favorable crystal structure \cite{Alijani_prb_11, Yamada_prm_18}. At the same time, a new possible disorder named L2$_1$B was reported to affect the SGS and the half-metallicity in quaternary Heusler alloys \cite{Ren_jsnm_16}. 
	
	The theoretical calculations suggest equi-atomic quaternary Heusler alloy such as CoFeVSi having a Y--type crystal structure shows a transition from SGS to half-metallic nature with the substitution of Mn atoms at V site \cite{Xu_epl_13, Ren_jsnm_16, Yamada_prb_19, Yamada_prm_18, Xiong_jmmm_14, Kharel_aip_22}. Recently, the spin band structure and density of states of CoFeVSi and CoFeMnSi Heusler alloys were examined considering the Y-type cyrstal structure \cite{Bainsla_prb_15, Yamada_prb_19, KudrnovskyPRB18}. A detailed analysis of magnetic behavior revealed a positive linear magneto-resistance in CoFeVSi, which induced changes in the minority spin band that resulted in a spin gapless structure \cite{Yamada_prm_18, Yamada_prb_19}. In a thin film of CoFeV$_{1-x}$Mn$_{x}$Si, a ferromagnetic behavior was reported consisting a Curie temperature ($T_{\rm C}$) varying between approximately 271~K (CoFeVSi) to around 620~K (CoFeMnSi) \cite{Yamada_prm_18, Bainsla_prb_15}. However, an experimental value of $T_{\rm C}=$ 657~K was reported for the CoFeV$_{0.5}$Mn$_{0.5}$Si Heusler alloy \cite{Kharel_aip_22}, which exceeded the value observed for the CoFeMnSi sample \cite{Bainsla_prb_15}. This makes the CoFeV$_{1-x}$Mn$_x$Si alloys interesting as the Mn substitution enhances the magnetic moment and Curie temperature \cite{Yamada_prm_18, Kharel_aip_22}. Also, the magnetic Heusler alloys are considered to be either perfect local systems or itinerant ferromagnetic systems, and their exchange couplings can be adequately characterized by a Heisenberg Hamiltonian \cite{Telling_prb_08, Rahman_prb_21, Shimizu_rpp_81}. However, several theoretical studies suggest that these Heusler alloys exhibit intricate exchange interactions of the localized magnetic moments \cite{Kurtulus_prb_05, Trudel_jap_10, Roy_prb_19}. The presence of long-range magnetism in these alloys is believed to play a crucial role in two types of couplings among neighboring spins: the direct exchange in the short-range coupling occurs between nearest--neighbor spins, and the long-range coupling between the spins mediated by the interaction of Ruderman-Kittel-Kasuya-Yosida (RKKY) type \cite{Trudel_jap_10}. 
	
More recently, the intrinsic critical behavior in Cobalt based ferromagnetic Heusler alloys is considered promising method to understand the magnetic interactions near the $T_{\rm C}$ \cite{Nehla_prb_19, Nehla_jap_19, Roy_prb_19, Rahman_prb_19, Rahman_prb_21}. The hypothesis of universality class states that certain signature of continuous phase transitions, such as critical exponents and scaling functions, are determined by global properties rather than microscopic details of the systems, such as space dimensionality, range of interaction, and order parameter symmetry~\cite{Stanley_rmp_99, Stanley_ox_71}. The analysis of data should be performed in the asymptotic critical region, defined as $\arrowvert$$\epsilon$$\arrowvert$~$\leq$ 10$^{-2}$, to obtain true asymptotic values of critical exponents and draw conclusions about singularity at the transition temperature \cite{Babu_jpcm_97, Kaul_jmmm_85}. Therefore, the critical behavior analysis in isotropic systems can be described using different universality classes near the $T_{\rm C}$ and the Neel temperature in ferromagnets and antiferromagnets, respectively \cite{Fischer_prb_02, Kaul_jmmm_85, Nehla_prb_19}. It has been well established that the values of critical exponents extracted from the detailed analysis of magnetic isotherms closely resemble the standard models \cite{Fisher_rpp_67, Kaul_jmmm_85, Guillou_prb_80, Fisher_prl_72, Fisher_prl_72, Campostrini_prb_02} such as mean field, 3D Heisenberg, 3D Ising, etc. On the other hand, experimental determination of the critical exponents are crucial to correlate with the universality classes and possibility of any deviation from established classical standard models \cite{Kim_prl_02, Padmanabhan_prb_07}. Also, the exchange interaction distance $J(r)$ was calculated to investigate magnetic ordering and spin interaction in full Heusler alloys \cite{Nehla_prb_19, Nehla_jap_19}. However, to the best of our knowledge these studies are still not available in detail for the quaternary Heusler alloys. 
	
Therefore, in this paper we report a detailed investigation of critical phenomena using magnetization measurements for the CoFeV$_{0.8}$Mn$_{0.2}$Si quaternary Heusler alloy across the paramagnetic to ferromagnetic transition temperature ($T_{\rm C}$ = 446$\pm1$~K). The Rietveld refinement of x-ray diffraction pattern confirms the single phase cubic Y--type structure with 28\% disorder between V and Si present in the sample. The magnetic susceptibility and critical behavior analysis reveal a second-order paramagnetic to ferromagnetic phase transition at around 446~K. More interestingly, the extracted values of critical exponents ($\beta$ = 0.369 and $\gamma$ = 1.445) suggest for the 3D Heisenberg type spin interactions in the sample. The Kouvel-Fisher approach and the Widom scaling relation are employed to validate the reliability of this investigation. Furthermore, the critical exponents indicate that the spin interaction decays as $J(r)$$\sim$r$^{(-4.99)}$, which suggests for the short-range 3D Heisenberg type magnetic ordering in the sample across the $T_{\rm C}$.

\section{\noindent ~Experimental}
	
A polycrystalline CoFeV$_{0.8}$Mn$_{0.2}$Si sample was synthesized using an arc melting system (CENTORR Vacuum Industries, USA) where high-purity constituent elements, namely Co, Fe, V, Mn, and Si, obtained from Alfa Aesar and/or Sigma Aldrich with a high purity 99.99\%, were melted in a water-cooled copper hearth. The melting process is done in the presence of a protective environment of dry argon gas and an additional 5 wt.\% of Mn was added to compensate for any evaporation losses due to its low vapor pressure. The weight loss during melting was observed below 0.9\%. The ingot sample was wrapped in Mo foil to prevent the reaction with the quartz tube at high temperatures and sealed in a quartz tube under a vacuum of approximately 10$^{-3}$ mbar. Subsequently, the sample underwent annealing at 1073~K for a duration of 14 days to enhance its homogeneity. The phase purity and crystal structure of the sample were analyzed using x-ray diffraction (XRD) measurements at room temperature using a PANalytical conventional x-ray diffractometer with a Cu K$\alpha$ ($\lambda$ = $1.5406$~\AA) source of radiation. The Rietveld refinement of XRD pattern was performed using the method implemented in the FULLPROF package. The sample composition was verified using energy-dispersive X-ray technique. The temperature and magnetic field-dependent magnetization data were recorded using a physical property measurement system (PPMS) from Quantum Design, USA. We use the low temperature system for the data in the temperature range of 2--300~K and magnetic field range of $\pm$3~Tesla, and the high temperature set-up for the data in the temperature range of 300--700~K and magnetic field range of 0--9~Tesla. 
	
\section{\noindent ~Results and discussion}

\begin{figure}[h]
\includegraphics[width=3.6in]{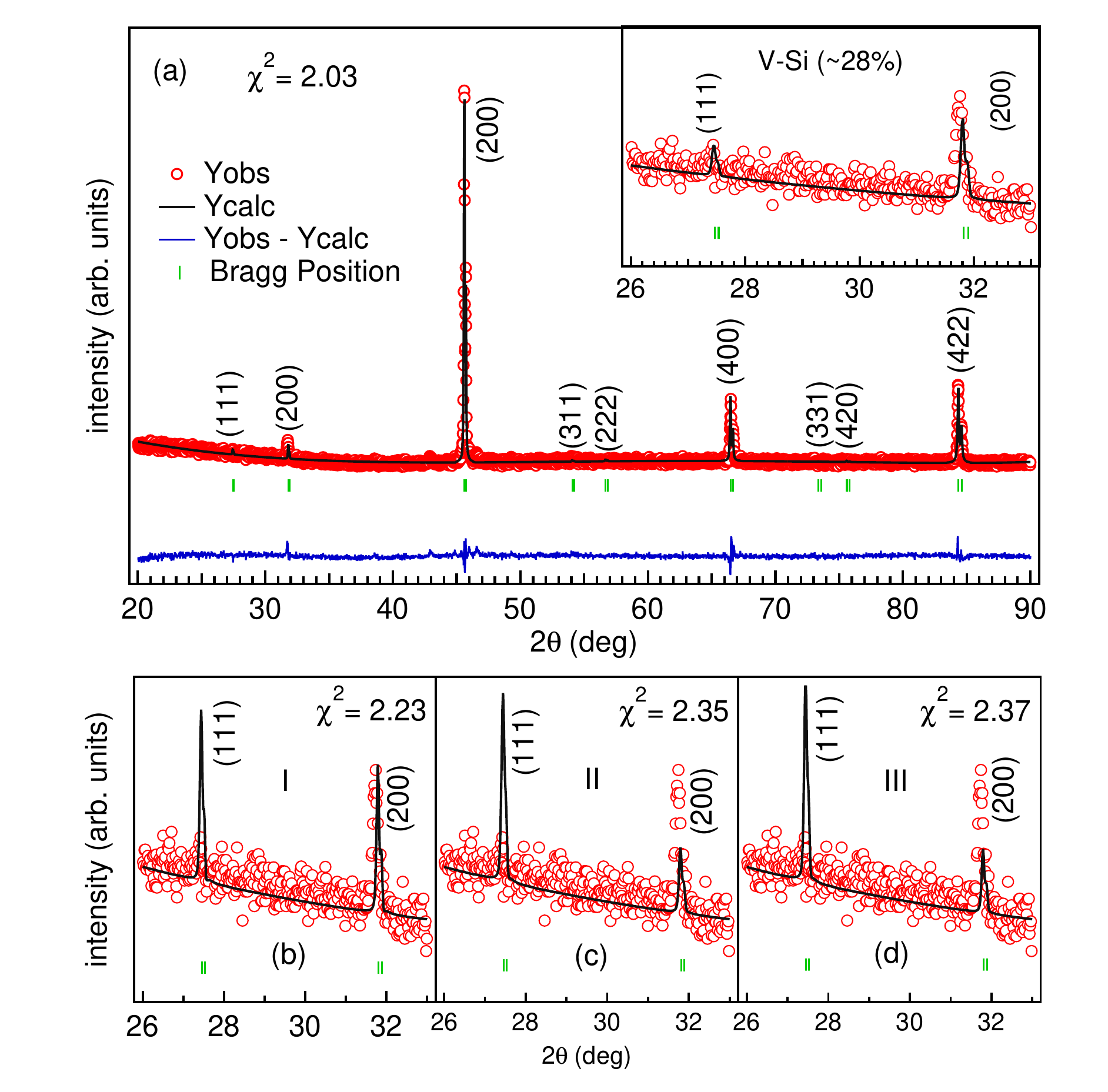}
\caption{(a) The Rietveld refined x-ray diffraction patterns of CoFeV$_{0.8}$Mn$_{0.2}$Si sample, the difference (solid blue line) between the observed (red open circle) and simulated (solid black line) patterns and Bragg positions (vertical green bars). The refinement of three different configurations of Y-type structure without anti-site disorder is shown in (b) for type--I, (c) for type--II, and (d) for type--III. The inset of (a) shows the fitting with the disorder between V and Si.} 
\label{Fig-Fig_1_XRD}
\end{figure}

The Rietveld refined x-ray diffraction (XRD) pattern of the CoFeV$_{0.8}$Mn$_{0.2}$Si sample, recorded in the 2$\theta$ angle range 20$^{\rm o}$ to 90$^{\rm o}$, is presented in Figs.~\ref{Fig-Fig_1_XRD}, which shows the single phase cubic Y--type structure with F$\bar{4}$3m (no. 216) space group \cite{Bainsla_prb_15, Venkateswara_prb_23}. The possible crystal structures were defined as Co, Fe, V/Mn and Si atoms randomly occupying Wyckoff positions 4d (3/4,3/4,3/4), 4c (1/4,1/4,1/4), 4b (1/2,1/2,1/2), and 4a (0,0,0) for all three configurations sites \cite{Alijani_prb_11, Nag_pra_23}. 
Note that when both the (111) and (200) superlattice peaks appear in the diffraction pattern, the Heusler alloy is regarded as a fully ordered $L2_1$ structure. However, these alloys often consist of disordered structures such as B2 and A2--types, where in the B2--type structure, (111) superlattice peak is absent, while both (111) and (200) superlattice peaks disappear in the A2--type structure \cite{Felser_sips_16, Graf_pssc_11}. We have performed the Rietveld refinement for the superlattice reflections (111) and (200) with three different configurations by using the corresponding atomic positions in the crystallographic axis. These refinements suggest the CoFeV$_{0.8}$Mn$_{0.2}$Si sample is stable in the type--I configuration, and the reliable parameter $\chi$$^2$ value of the fitting is the lowest for type--I rather than other configurations such as type--II and type--III, as presented in Figs.~\ref{Fig-Fig_1_XRD}(b-d). The theoretical calculations suggest the disorder V--Fe and Co--Fe atoms, and the structure is so-called the L2$_1$B--type structure \cite{Ren_jsnm_16}; however, it is difficult to find the disorder between Co--Fe and V--Fe with conventional XRD pattern because these elements have similar x-ray scattering cross-section \cite{Ren_jsnm_16, Yamada_prb_19}. In order to find the best fit, we modeled the refinement considering the anti-site disorder between V and Si due to the significant difference in the scattering cross section \cite{Felser_sips_16}. In Fig.~\ref{Fig-Fig_1_XRD}(a), the XRD pattern of the sample belongs to the most stable type--I, where the Wyckoff position of the atom is assigned as Co at 4d, Fe at 4c, V/Mn at 4b, and Si at 4a~\cite{Alijani_prb_11}. In our sample, the estimated anti-site disorder is $\approx$28\% between V--Si where the atoms can exchange their position $\approx$28\% of the total occupancy in the stable type--I. The reliable parameter $\chi$$^2$ improved from 2.23 to 2.03 after considering the anti-site disorder, resulting in the best fit is presented in inset of Fig.~\ref{Fig-Fig_1_XRD}(a). The obtained lattice parameter 5.621~\AA~is good agrees with the refs.~\cite{Yamada_prb_19, Ozdogan_jap_13}. Further, we have investigated the composition of the sample and found to be $\approx$27:24:21:4:24, which is close to the stoichiometry ratio.
		
\begin{figure}[ht]
\includegraphics[width=3.5in]{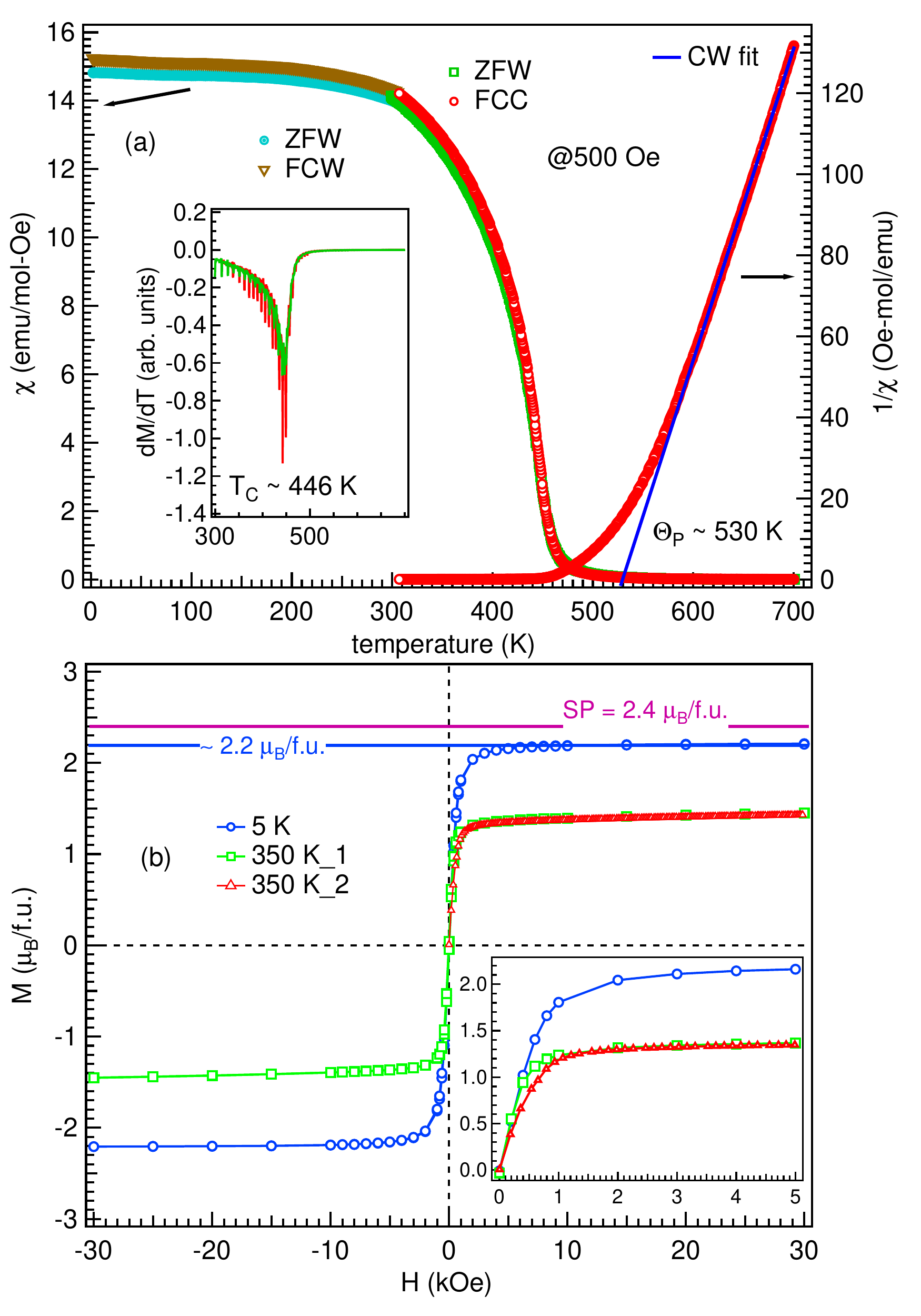}
\caption{(a) The temperature dependent magnetic susceptibility ($\chi$) recorded at 500~Oe magnetic field along with the derivative of $M$ in the inset. On the right axis, the $\chi^{-1}$ is plotted and fitted (solid blue line) using the Curie-Weiss law between 610--700~K and extrapolated up to zero. (b) The magnetic isotherms ($M-H$) measured at 5~K, and the data at 350~K are matching well from two different systems used.}
\label{Fig-Fig_2_MT_MH}
\end{figure}

The dc-susceptibility ($\chi$ versus temperature) data are measured at applied magnetic field of 500~Oe in both the modes zero--field--warming (ZFW) and field-cooled-warming (FCW) in the temperature range of 2--300~K and ZFW and field--cooled--cooling (FCC) in 300--700~K, as shown on the left axis of Fig.~\ref{Fig-Fig_2_MT_MH}(a). As we use two different instruments for low and high temperature range, there is a slight mismatch in the magnetization values at 300~K; therefore, the low temperature data are shifted upwards by 2~emu/mol--Oe. Here, we find that there is no distinct bifurcation between the ZFW and FCC curves at low temperatures, which suggests for the ordered magnetism, i.e., there is no magnetic frustration in this sample. In order to find the paramagnetic (PM) to ferromagnetic (FM) transition temperature ($T_{\rm C}$), we plot the derivative of magnetization (dM/dT) as a function of temperature in the inset of Fig.~\ref{Fig-Fig_2_MT_MH}(a), which clearly confirms the $T_{\rm C}$ value of $446$~K. 
Furthermore, we plot the inverse of susceptibility and used the Curie--Weiss (C--W) law $\chi$ = C/(T -- $\Theta_{\rm P}$) to establish a linear relation (solid blue line) in high-temperature range (610--700~K) to estimate the C--W temperature ($\Theta_{\rm P}$), as shown in Fig.~\ref{Fig-Fig_2_MT_MH}(a) on the right axis. The extracted value of $\mu$$_{eff}$ is 3.2~$\mu$$_{\rm B}$/f.u.~and $\Theta_{\rm P}$ is determined to be about $530$~K, which is greater than the $T_{\rm C}$ value due to persistence of ferromagnetic interactions above the transition temperature \cite{Fischer_prb_02}. 
\begin{figure}
\includegraphics[width=3.55in]{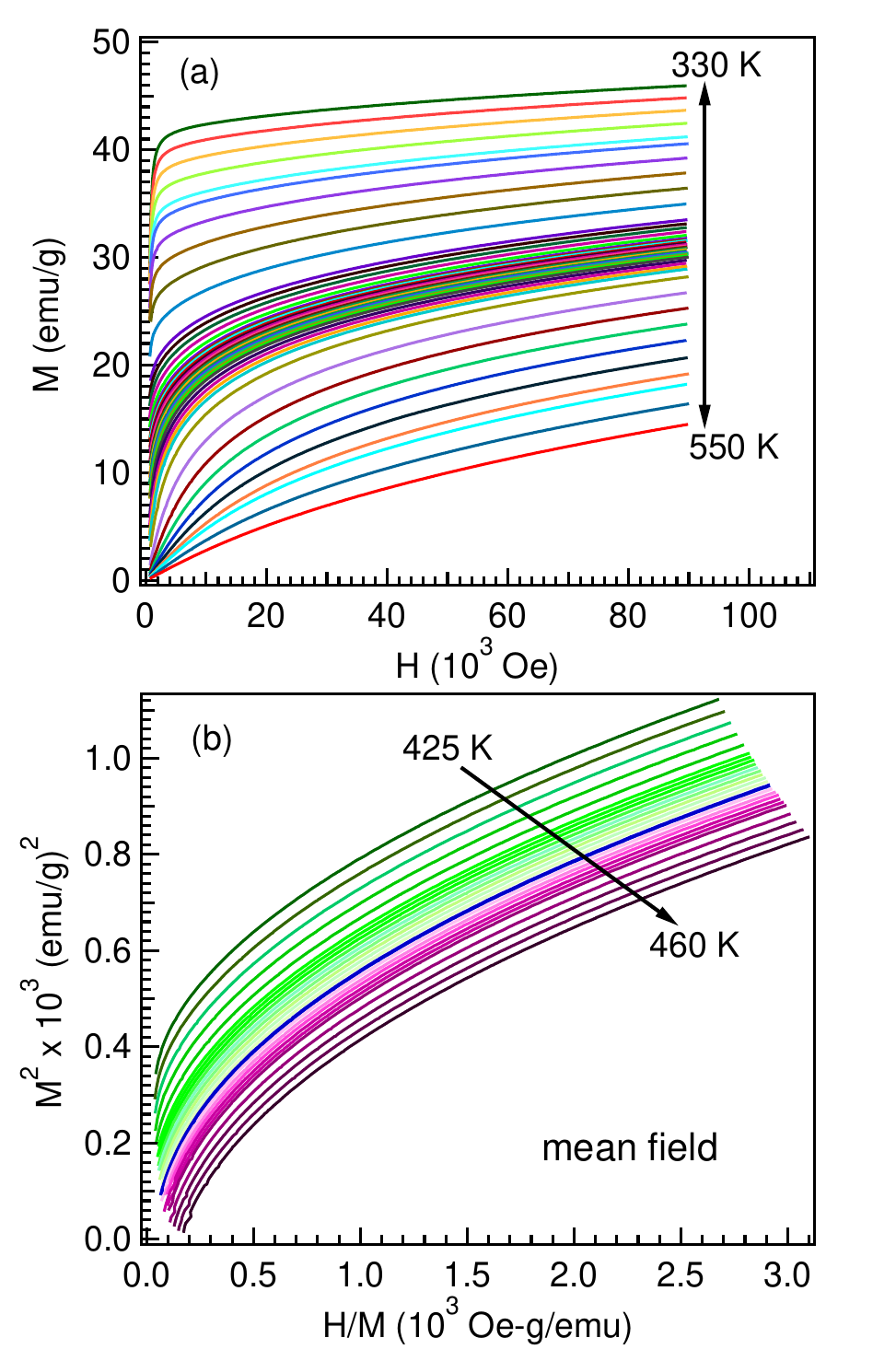}
\caption{(a) The isothermal magnetization ($M-H$) curves are recorded in the temperature range of 330--550~K having 1~K step in 440--452~K range across the transition temperature ($T_{\rm C}$). (b) The Arrott plot ($M^2$ versus $H/M$) in the temperature range of 425--460~K where the blue curve corresponds to the $T_{\rm C}=$ 446~K.} 
\label{Fig-Fig_4_MH_mean_field}
\end{figure}
Fig.~\ref{Fig-Fig_2_MT_MH}(b) presents the magnetic isotherms ($M-H$) at 5~K and 350~K temperature in the magnetic field range of $\pm$3~Tesla. To check the consistency, we have recorded the $M-H$ data using two different instruments at 350~K using low temperature module (350 K$_1$, green color) and high temperature set-up (350 K$_2$, red color), which are found to be overlap with each other. The absence of a hysteresis in an increasing and decreasing legs of magnetic field indicates the soft ferromagnetic nature of the sample. Although it is reported that the coercive field increases with an increase in the concentration of Mn element at V sites at 300~K \cite{Yamada_prb_19}. The saturation magnetization are turn out to be 2.2~$\mu$$_{\rm B}$/f.u.~and 1.4~$\mu$$_{\rm B}$/f.u.~at 5~K and 350~K temperature, respectively. The experimental value of saturation magnetization was reported 1.5~$\mu$$_{\rm B}$/f.u. for 300~K in CoFeV$_{0.8}$Mn$_{0.2}$Si thin film \cite{Yamada_prb_19, Yamada_prm_18} and thus in our sample the saturation value approximately consistent with both the Y-type and L2$_1$B type structure, which has been theoretically calculated in Ref.~\cite{Ren_jsnm_16}. The half-metallic ferromagnetic Heusler alloys follow the Slater-Pauling (S--P) rule, given by M$_{\rm t}$ = (Z$_{\rm t}$--24)~$\mu$$_{\rm B}$/f.u., where M$_{\rm t}$ represents the overall magnetic moment and Z$_{\rm t}$ denotes the total count of valence electrons within the unit cell \cite{Felser_sips_16}. Here, the saturation magnetization experimentally observed at 5~K temperature (solid blue line) concurs well with the value of $M_S$ (2.4~$\mu$$_{\rm B}$/f.u.) from the S--P rule and is shown with the solid magenta line in Fig.~\ref{Fig-Fig_2_MT_MH}(b) \cite{Galanakis_prb_02}. Moreover, the isothermal virgin ($M-H$) curves around the ferromagnetic transition temperature ($T_{\rm C}$) are recorded in the range of 330--550~K up to 9~T external applied magnetic field in the high temperature set-up, as shown in Fig.~\ref{Fig-Fig_4_MH_mean_field}(a). The $M-H$ curve at a high-temperature (550~K) shows the non-linear behavior, owing to the formation of ferromagnetic cluster significantly higher than $T_{\rm C}$, a similar behavior was reported up to 2$T_{\rm C}$ in Ref.~\cite{Fischer_prb_02}. Further, we present the Arrott plot between $M^2$ and $H/M$ in Fig.~\ref{Fig-Fig_4_MH_mean_field}(b), discussed later with detailed analysis. 
	  
\begin{figure}
\includegraphics[width=3.55in]{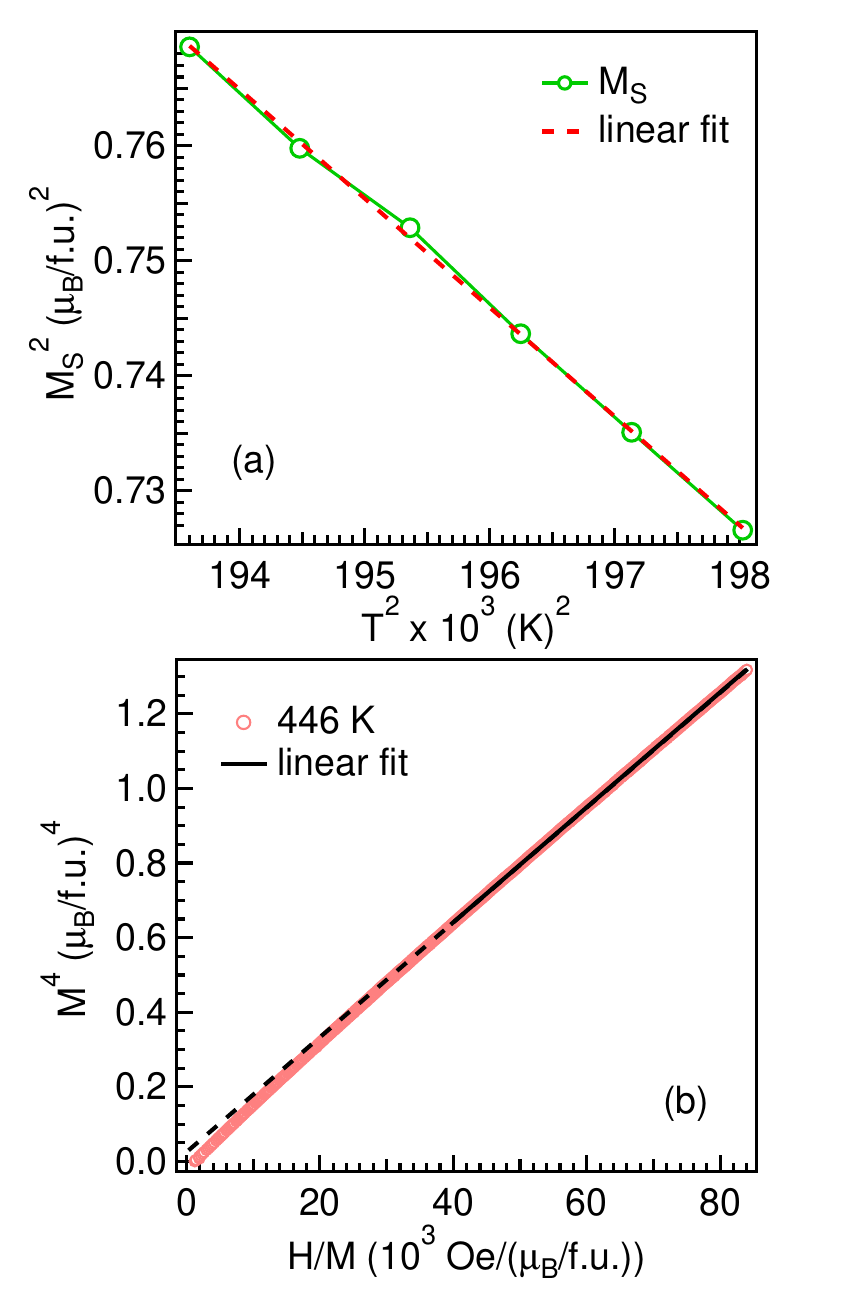}
\caption{(a) The $M_{\rm S}$$^2$ versus $T^2$ below $T_{\rm C}$ and dotted red line is the linear fit. (b) The $M^4$ versus $H/M$ of isotherms at $T_{\rm C}$ and solid black line represents linear fit in the high magnetic field regime and dotted line is extrapolated.}
\label{Fig-Fig_3_SCR}
\end{figure}
	
First, in order to gain insights into the nature of magnetism whether localized or itinerant \cite{Qin_prb_20}, we employed the Rhodes-Wohlfarth ratio (RWR) \cite{Rhodes_prsla_63, Saunders_prb_20,Gupta_prb_22}, which is defined as the ratio of the paramagnetic moment (P$_{\rm C}$) to the saturation magnetization (P$_{\rm S}$) at low temperature (5~K), which is considered as the ground state as 5~K is much less than the $T_{\rm C}$ \cite{Shambhu_prm_19}. The value of P$_{\rm C}$ is calculated using the formula: $\mu_{eff}$ = $\sqrt{\rm P_{\rm C}(\rm P_{\rm C} + 2)}$, where $\mu_{eff}$ is derived from the C--W fitting of the magnetization data and used to determine P$_{\rm C}$. A RWR value close to 1 suggests the presence of localized moments, while a value greater than 1 indicates a more traditional itinerant system \cite{Saunders_prb_20, Fischer_prb_02}. In our sample, RWR was found to be 1.1, which is relatively low compared to other conventional itinerant ferromagnetic materials described in the literature \cite{Saunders_prb_20, Bhattacharyya_prb_11, Mondal_prb_21}, such as FeMnVAl with a RWR of 1.51 \cite{Gupta_prb_22}. However, in our case, the RWR is close to unity, which is expected for itinerant magnetic systems. Furthermore, the self-consistent renormalization (SCR) theory, provides a comprehensive framework for studying itinerant electron systems \cite{Takahashi_jpsj_86, Mondal_prb_21}. This theory incorporates the conservation of both thermal and zero-point spin fluctuations, as reported in refs.~\cite{Takahashi_jpsj_86, Mondal_prb_21}. Notably, the magnetization versus magnetic field ($M-H$) curves in the vicinity of the Curie temperature (T$_{\rm C}$) is primarily influenced by the thermal spin fluctuation \cite{Saunders_prb_20}. To analyze this behavior, we plot the $M_{S}$$^2$ versus $T^2$ below $T_{\rm C}$ in Fig.~\ref{Fig-Fig_3_SCR}(a), and the $M^4$ versus (H/M) in Fig.~\ref{Fig-Fig_3_SCR}(b), at critical isotherms, which are expected to follow linear behavior. However, in our case it shows a small deviation from linearity as the linearity is visible only in the large field region, as indicated in Fig.~\ref{Fig-Fig_3_SCR}(b), which suggests for the non-dominant character of itinerant magnetism in the sample \cite{Moriya_jpsj_78, Takahashi_jpsj_86}. To quantify this effect, the SCR theory suggests that $M-H$ curve at $T_{\rm C}$ follows the Eq.~\ref{Eq:Eq_1_SCR}, as given below~\cite{Mondal_prb_21}:
\begin{equation}
M^4=\frac{1}{4.671}\left(\frac{T_{\rm C}^2}{T_A^3}\right)\left(\frac{H}{M}\right),
\label{Eq:Eq_1_SCR}
\end{equation}
where $M$ and $H$ quantities are expressed in the units of $\mu_{\rm B}$/f.u. and Oe, respectively. The parameter $T_A$ represents the dispersion of the spin fluctuation spectrum in wave-vector space, measured in Kelvin (K). A linear relationship between magnetization and temperature ($T_{\rm C}$) is commonly observed in many itinerant ferromagnetic alloys, as reported in refs.~\cite{Mondal_prb_21, Gupta_prb_22}. However, in the present case, the curve deviates from linearity indicating that the magnetism in this material is not only primarily governed by itinerant behavior, but may instead exhibit characteristics of localized Mn-spin moments \cite{Gupta_prb_22, Mondal_prb_21, Qin_prb_20}. The linear fit of the data using the Eq.~\ref{Eq:Eq_1_SCR} yields a slope of the line equal to 1.578(1)$\times$10$^{-5}$ ($\mu_{\rm B}$/f.u.)$^5$/Oe. Using the obtained value of $T_{\rm C}$ and the slope, we calculate the value of $T_A$ to be 1392~K. Additionally, the expression for the $T_{\rm C}$ is governed by SCR theory, as described in ref.~\cite{Mondal_prb_21}, is given by the equation $T_{\rm C}$ = (60c)$^{-3/4}T_A^{3/4}P_S^{3/2}T_0^{1/4}$, where $c$ is a constant term with a value of 0.3353 \cite{Mondal_prb_21}. By using this equation and the values of $T_A$ and $P_S$, we can determine the width of the dynamical spin fluctuation ($T_0$) to be around 1053.5~K. According to the SCR theory, the degree of localization or itinerancy of the spin moment is determined by the ratio $T_{\rm C}/T_0$. When $T_{\rm C}/T_0$ is much smaller than 1, magnetic materials exhibit strong itinerant character. On the other hand, if $T_{\rm C}/T_0$ is approximately equal to 1, the materials show local moment magnetism \cite{Takahashi_springer_13}. In our sample, the calculated ratio of $T_{\rm C}/T_0$ found to be 0.42, indicating itinerant ferromagnetic behavior \cite{Mondal_prb_21, Takahashi_springer_13, Babu_jpcm_97}. 
	
\begin{figure*}
\includegraphics[width=7.3in]{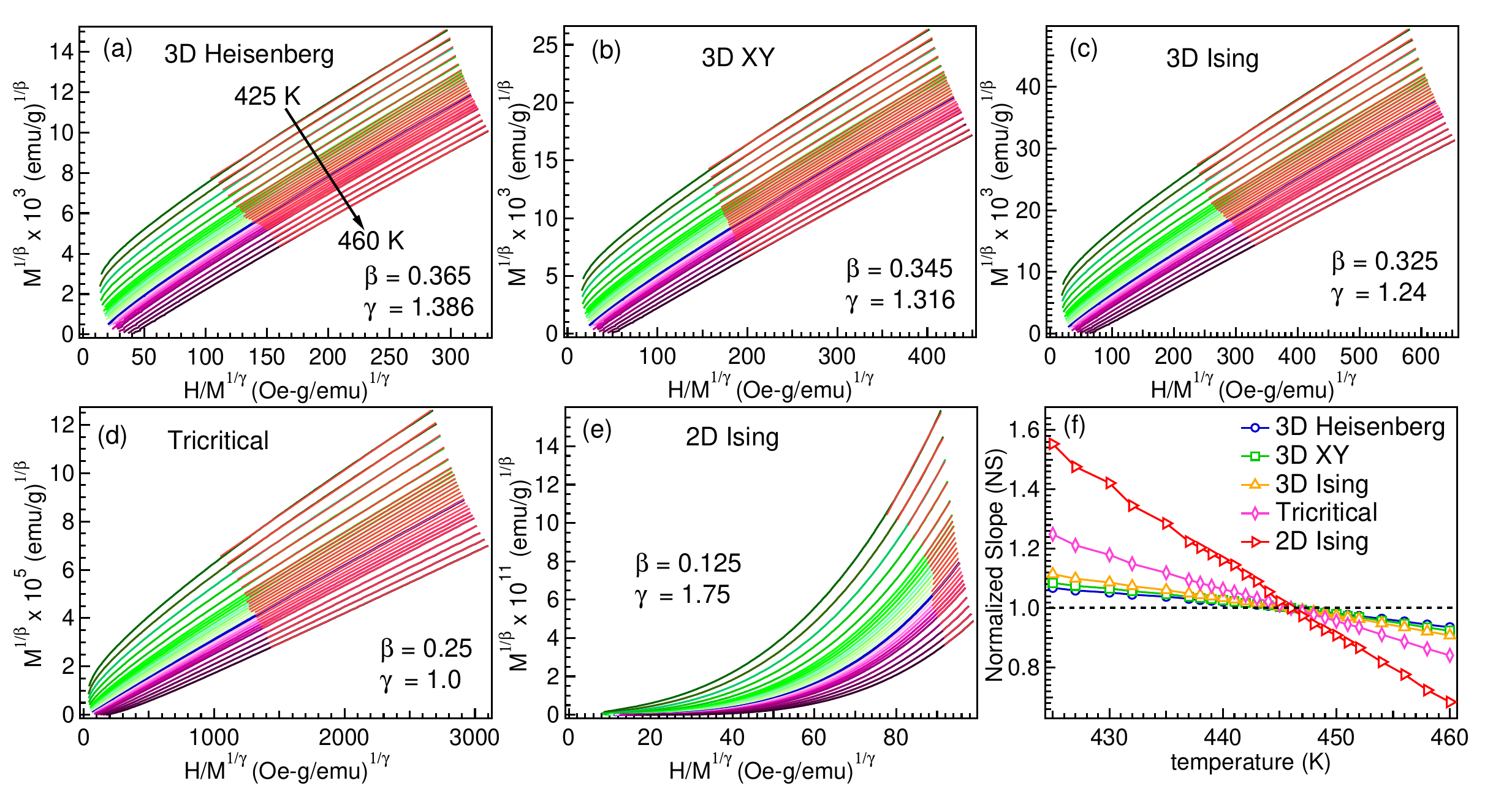}
\caption {The isotherms plots $M^{1/\beta}$ versus $H/M^{1/\gamma}$ in the temperature range of 425--460~K using the (a) 3D Heisenberg, (b) 3D XY, (c) 3D Ising, (d) Tricritical mean field, and (e) 2D Ising models. (f) The normalized slope; NS = S($T$)/S($T_{\rm C}$) determined by the linear fitting of data in (a--e) at each temperature.}
\label{Fig-Fig_5_H_XY_Ising}
\end{figure*}
	
Now we move back to perform the detailed analysis of isothermal magnetization to understand the magnetic interactions across the $T_{\rm C}$ by extracting the values of critical exponents ($\beta$, $\gamma$, and $\delta$). For this purpose, the magnetic isotherm ($M-H$) curves measured in the temperature range of 440--452~K with 1~K step across the $T_{\rm C}$ are shown in Fig.~\ref{Fig-Fig_4_MH_mean_field}(a), where the asymptotic region is defined as $|\epsilon|$\textless0.01 according to the criteria mentioned in the refs.~\cite{Babu_jpcm_97, Kaul_jmmm_85}. We observe a steep rise in the magnetization at low fields and non-saturating behavior in the high field range up to 9~T, which indicates the itinerant behavior of ferromagnetism in this sample \cite{Kaul_jpcm_99, Bhattacharyya_prb_11}. For any spin interaction system, the transition from paramagnetic (PM) to ferromagnetic (FM) behavior can be determined by constructing a plot between M$^{1/\beta}$ and (H/M)$^{1/\gamma}$. According to the Landau mean-field theory, the conventional Arrott plot \cite{Arrott_prl_67} exhibits a linear relationship representing the long-range ferromagnetic interaction near the critical temperature $T_{\rm C}$, with critical exponents $\beta = 0.5$ and $\gamma = 1$. The resulting plot should consist of multiple parallel straight lines, with one line passing through the origin \cite{Babu_jpcm_97, Nehla_prb_19}. The slope of these lines is related to the system's critical exponent $\beta$, which measures how magnetization varies with the temperature in the proximity of the transition temperature. To construct an Arrott plot, the quantity $(M^2)$ is plotted in Fig.~\ref{Fig-Fig_4_MH_mean_field}(b) as a function of $H/M$ for various temperature values across the $T_{\rm C}$. These curves show a substantial curvature, which indicates the deviation from the Landau mean-field theory and simply extrapolation approach may give rise to large uncertainty \cite{Kaul_jmmm_85, Babu_jpcm_97}. This non-linear behavior of the Arrott plot suggests that the itinerant ferromagnetism within the framework of the conventional Landau mean-field model, which should possess the spin fluctuation and electronic correlation, can be neglected in the present case \cite{Roy_prb_19, Fischer_prb_02}. Therefore, we need to perform detailed critical analysis to understand the magnetic interactions across $T_{\rm C}$ using Banerajee's criteria \cite{Banerjee_pl_64} where the magnetic equation of state specifies the second-order magnetic phase transition from PM to FM by employing the critical exponents $\alpha$, $\beta$, and $\gamma$, which are all mutually related to the magnetic behavior of the sample. The values of these critical exponents can be utilized thoroughly to investigate the magnetic interactions across the second-order phase transition \cite{Banerjee_pl_64}. 

\begin{figure}
\includegraphics[width=3.4in]{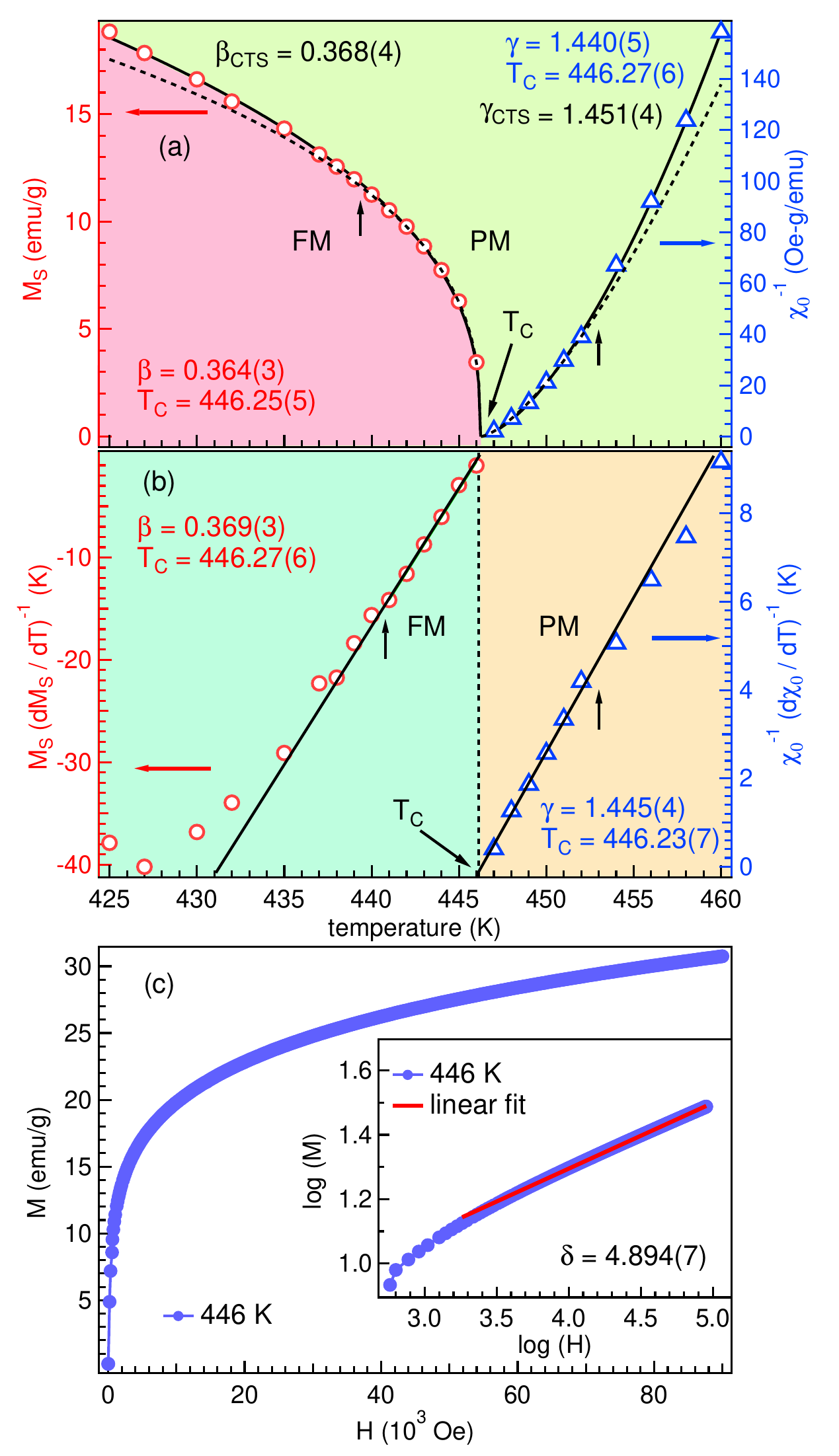}
\caption {(a) The spontaneous magnetization $M_S$ (red horizontal arrow) and inverse initial susceptibility $\chi_0$$^{-1}(T)$ (blue horizontal arrow), fitted using the equations~(\ref{Eq:Eq_3_beta}) and~(\ref{Eq:Eq_4_gamma}) (dashed lines), and using equations~(6) and (7) (solid lines), respectively. (b) The Kouvel-Fisher plot  $M_S/(dM_S/dT)^{-1}$ (red horizontal arrow) and $\chi_0^{-1}$/($d\chi_0^{-1}/dT)^{-1}$ (blue horizontal arrow) as a function of temperature below and above $T_{\rm C}$, respectively. The solid black lines are the straight line fit using equations~8 and 9. The black vertical arrows in (a, b) indicate the deviation in the far temperature range. (c) The Isothermal magnetization is at $T_{\rm C}=$ 446~K and inset is the same plot in $\log (M)$-$\log (H)$ scale where the high field region is fitted with a straight line (solid red line) using equation~(\ref{Eq:Eq_5_delta_TC}).}
\label{Fig-Fig_6_Ms_beta_gamma}
\end{figure} 
	
To investigate the universal scaling laws associated with the spontaneous magnetization $M_S$ and initial susceptibility $\chi_0^{-1}$ near a second-order phase transition, the divergence of the correlation length is expressed as $\xi = \xi_0|(T-T_{\rm C}/T_{\rm C})|^{-\nu}$~\cite{Zhang_prb_12,Nehla_jap_19}. Here, $\xi$ represents the correlation length, $\xi_0$ is a constant, $T_{\rm C}$ denotes the Curie temperature and the exponent $\nu$ characterizes the critical behavior of the system. Here, we speculate the magnetization isotherms by calculating the true critical exponents using the modified Arrott plots (MAPs). Therefore, the Arrott-Nokes equation of state \cite{Arrott_prl_67} in the asymptotic region $|\epsilon|\leq 0.01$ is given as follows:
\begin{equation}
(H/M)^{1/\gamma}=A\epsilon+B M^{1/\beta}
\label{Eq:Eq_2_Arrott-Nokes}
\end{equation}	
where A and B are constant terms and $\epsilon$ is the reduced temperature, which is defined as $\epsilon$ = ($T-T_{\rm C}$)/$T_{\rm C}$, and $\beta$, $\gamma$ are the critical exponents. We performed a strict iterative process with proper values of $\beta$, $\gamma$ and $T_{\rm C}$) to generate the set of parallel lines and to eliminate nonphysical and systematic errors in exponent values. The intercept on the $M^{1/\beta}$ and $(H/M)^{1/\gamma}$ axes is obtained via linear extrapolation of high field region, yielding the starting values of $M_S(T)$ and $\chi_0$$^{-1}(T)$, which are utilized to generate the MAPs by using equation (\ref{Eq:Eq_2_Arrott-Nokes}). Although there are two approaches to find the best model for correct $\beta$ and $\gamma$ values: (1) the straight line must be present in the high field region, and (2) the slope of the line must be the same, which implies that these must be parallel in the MAPs. In Figs.~\ref{Fig-Fig_5_H_XY_Ising}(a--e), we present tha constructed MAPs using the critical exponents from standard model foretold for the three-dimensional (3D) systems such as 3D Heisenberg ($\beta$ = 0.365, $\gamma$ = 1.386), 3D XY ($\beta$ = 0.345, $\gamma$ = 1.316), 3D Ising ($\beta$ = 0.325, $\gamma$ = 1.24), tricritical mean field ($\beta$ = 0.25, $\gamma$ = 1) and 2D Ising ($\beta$ = 0.125, $\gamma$ = 1.75) models \cite{Guillou_prb_80, Stanley_ox_71, Fisher_rpp_67, Kim_prl_02}. We find that the lines in the high field regions are not straight and not parallel to each other; therefore, as the second criterion we calculate the normalized slope (NS) at each temperature near the $T_{\rm C}$ using the expression: NS = S$(T)$/S($T_{\rm C}$) \cite{Nehla_prb_19,Roy_prb_19}, where the slope of $M^{1/\beta}$ versus $(H/M)^{1/\gamma}$ at a temperature $T$ is denoted by S($T$), and the S($T_{\rm C}$) is the slope at Curie temperature. The NS value must follow the unity for the best fit universality model and following this criteria  the plots in Fig.~\ref{Fig-Fig_5_H_XY_Ising}(f) clearly indicate that the 3D Heisenberg model follows minimum deviation from the unity below and above the $T_{\rm C}$. Thus, it can be concluded at this stage that the 3D Heisenberg type interactions better defines the critical behavior across the $T_{\rm C}$ in the sample. Further, to confirm the applicability of this model, which reveals the ordering near $T_{\rm C}$ by calculating the $\beta$ (for $M_S$), $\gamma$ (for $\chi_0^{-1}$), and $\delta$ (magnetization at critical isotherms) in the asymptotic region using the following simple power-law of the magnetic equation of state \cite{Fisher_rpp_67, Kouvel_pr_64, KouvelPRL68}. 	
\begin{equation}
M_{\mathrm{S}}(0, T)=M_0(-\epsilon)^\beta, \quad \epsilon<0,~T<T_{{C}},
\label{Eq:Eq_3_beta}
\end{equation}
\begin{equation}
\chi_0^{-1} (0,T)=\frac{h_0}{M_0}(\epsilon)^\gamma, \quad \epsilon>0,~T>T_{{\rm C}},
\label{Eq:Eq_4_gamma}
\end{equation}
\begin{equation}
{M (H, {T}_{C})}=\mathrm{DH}^{1 / \delta}, ~~~~~~ \epsilon=0, {~T}={T}_{{\rm C}}
\label{Eq:Eq_5_delta_TC}
\end{equation}
where $\epsilon$ is called the reduced temperature, and the critical amplitudes are M$_0$, $h_0/M_0$, and D. The equations 3 and 4 are applicable only when the temperatures are very close to the $T_{\rm C}$; means in the limit when $\arrowvert$$\epsilon$$\arrowvert$ tends to zero \cite{Babu_jpcm_97}. 
		
Note that the $M_S$ and $\chi_0^{-1}$ values can be used to calculate the critical exponents $\beta$ and $\gamma$ by fitting the curves using equations 3 and 4 over the temperature range $\arrowvert$$\epsilon$$\arrowvert$$\leq$0.01. Subsequently, the extracted values of $\beta$ and $\gamma$ are employed to reconstruct the MAPs and the straight lines are extrapolated in the high field region, then the new values of $M_S$ and $\chi_0^{-1}$ are obtained from the plots. We use the optimal theoretical model and repeat the iteration method until the convergent of new $\beta$ and $\gamma$ values having a better fit to the data. 
\begin{table*}[ht]
\caption{The extracted values of critical exponents ($\alpha$, $\beta$, $\gamma$, and $\delta$) from the modified Arrott plot (MAP), Kouvel-Fisher (KF) technique, correction-to-scaling (CTS) and critical isotherm (CI) for CoFeV$_{0.8}$Mn$_{0.2}$Si Heusler alloy and comparison with the theoretically anticipated values for different universality classes.}
			\begin{tabular}{p{2.7cm}p{2.8cm}p{1.7cm}p{1.7cm}p{1.4cm}p{1.4cm}p{1.4cm}p{1cm}p{1cm}p{1.2cm}}
		\hline
		\hline
		Sample & Reference & Method(s) &$T_C$& $\alpha$& $\beta$& $\gamma$&$\delta$&$a_{M_2}^{-}$&$a_{\chi_2}^{+}$\\
		\hline
		CoFeV$_{0.8}$Mn$_{0.2}$Si& This work &MAP& 446.25(5)& --0.168(3)&0.364(3)&1.440(6)&4.96&&\\
		
		& This work &CTS &446.25(5)& &0.368(4) & & &0.53(6)&\\
		
		& This work &CTS &446.27(6)& & & 1.451(4)& &&--0.91(4)\\
		
		& This work &KF&446.23(7)& --0.183(2)&0.369(3)&1.445(4)&4.92&&\\
		
		& This work &CI&  & & &&4.89&&\\
		
		Mean field & \cite{Kaul_jmmm_85} and \cite{Stanley_ox_71} &Theoretical && ~0 & 0.5& 1.0& 3.0 &&\\
		3D Heisenberg & \cite{Fisher_rpp_67}, \cite{Stanley_ox_71} and \cite{Guillou_prb_80} &Theoretical && --0.116 & 0.365 & 1.386&4.80 &&\\
		3D XY &\cite{Fisher_rpp_67}, \cite{Stanley_ox_71} and \cite{Guillou_prb_80} &Theoretical  && --0.006 & 0.345 & 1.316& 4.82&&\\
		3D Ising & \cite{Fisher_rpp_67}, \cite{Stanley_ox_71} and \cite{Guillou_prb_80} &Theoretical  && 0.009 & 0.325 & 1.24& 4.81 &&\\
		Tricritical & \cite{Kim_prl_02} & Theoretical && 0.5&0.25 & 1.0 & 5.0 && \\
		\hline
		\hline 
		\label{tab: critical exponents}
	\end{tabular}
		\end{table*}
The estimation of critical exponents using this approach is independent of the original parameters, confirming the analysis's validity and the intrinsic character of the derived critical exponents. After a few iterations, the $M_S$ and $\chi_0^{-1}$ estimated from the linear extrapolation in the high field region up to zero field, which gives the positive intercept ($M_S$ values) on the $y-$axis $(M)^{1/\beta}$ and also positive intercept ($\chi_0^{-1}$ values) on the $x-$axis $(H/M)^{1/\gamma}$. The obtained values of the $M_S$ and $\chi_0$$^{-1}$ are displayed as a function of temperature in Fig.~\ref{Fig-Fig_6_Ms_beta_gamma}(a). In practical scenarios, extrapolation becomes necessary for numerous systems due to the deviation of isotherms from a straight-line behavior at very low fields, making it imperative to rely on the bahavior in high-field region \cite{Fischer_prb_02}. By fitting the values in Fig.~\ref{Fig-Fig_6_Ms_beta_gamma}(a), we find the stable values of $\beta$ (0.364) and $\gamma$ (1.440) with the $T_{\rm C}$ ($\sim$446~K), as shown by the black dotted lines where the vertical arrows indicate the deviation at the temperature values far away from the $T_{\rm C}$. In practice for most of the cases, the critical exponents are extracted using simple power-law fits in a very wide temperature range across the $T_{\rm C}$; however, the effective values may depend on the chosen temperature range. Therefore, here we perform the analysis with correction-to-scaling (CTS) terms to find the critical exponents in the wide range of temperature (--0.04\textless$\epsilon$\textless0.03) across the $T_{\rm C}$, as suggested in ref.~\cite{Babu_jpcm_97}. In this case, the following modified magnetic equation of state are used:
\begin{equation}
		\begin{array}{ll}
			M_{\mathrm{S}}(0, T)=M_0(-\epsilon)^\beta\left[1+a_{M_1}^{-}|\epsilon|^{\Delta_1}+a_{M_2}^{-}|\epsilon|^{\Delta_2}\right] & \epsilon<0 
		\end{array}
	\end{equation}
	\begin{equation}
		\begin{array}{ll}
			\chi_0^{-1} (0,T) =\left(h_0 / M_0\right) \epsilon^{\gamma}\left[1+a_{\chi_1}^{+}|\epsilon|^{\Delta_1}+a_{\chi_2}^{+}|\epsilon|^{\Delta_2}\right]^{-1} & \epsilon>0
		\end{array}
	\end{equation}
where ($a_{M_1}^{-}$, $a_{M_2}^{-}$) and ($a_{\chi_1}^{+}$, $a_{\chi_2}^{+}$) are the amplitudes, $\Delta_1$ and $\Delta_2$ are the leading CTS critical exponents, while remaining terms have the same meaning as aforementioned. Here, only the leading CTS terms $\Delta_2$ is considered during fitting because $\Delta_1$ (\textless \textless $\Delta_2$) can be neglected in the wide temperature range away from the $T_{\rm C}$. Using these equations \cite{Babu_jpcm_97}, the $M_S(T, 0)$ and $\chi_0^{-1} (0,T)$ data in Fig.~\ref{Fig-Fig_6_Ms_beta_gamma}(a) are fitted (as shown by the solid black lines) in the temperature range of 425--460~K. The obtained value of critical exponents using CTS terms are found to be very similar to the values extracted from simple-power-law fit in the asymptotic region of the $T_{\rm C}$, as summarized in Table~I. 

Furthermore, the Kouvel-Fisher method (KF) \cite{Kouvel_pr_64} is used to validate the accuracy of the values of the critical exponents extracted from the MAP, where equations~(\ref{Eq:Eq_6_Kouvel}) and (\ref{Eq:Eq_7_Fisher}) are modified according to the Kouvel-Fisher method, as below: 
		\begin{equation}
			\frac{M_{\mathrm{S}}(T)}{d [M_{\mathrm{S}}(T) / dT]}=\frac{T-T_{\mathrm{C}}}{\beta}, 
			\label{Eq:Eq_6_Kouvel}
		\end{equation}	
		\begin{equation}
			\frac{\chi_0^{-1}(T)}{d [\chi_0^{-1}(T) / dT]}=\frac{T-T_{\mathrm{C}}}{\gamma},
			\label{Eq:Eq_7_Fisher}
		\end{equation}
The value of $T_{\rm C}$ is determined by the intersection of $M_S/(dM_S/dT)^{-1}$ and $\chi_0^{-1}$/($d\chi_0^{-1}/dT)^{-1}$ on the temperature axis, as shown in Fig.~\ref{Fig-Fig_6_Ms_beta_gamma}(b), which should both be straight lines with slopes of $\frac{1}{\beta}$ and $\frac{1}{\gamma}$, respectively, as determined by above equations~(\ref{Eq:Eq_6_Kouvel}) and (\ref{Eq:Eq_7_Fisher}). The linear fit to the data in Fig.~\ref{Fig-Fig_6_Ms_beta_gamma}(b) give rise the values of $\beta$, $\gamma$ and T$_{\rm C}$, which are found to be 0.369(3), 1.445(4) and 446~K, respectively. It is important to note here that the values of critical exponents found to be similar in both the cases (magnetic equation of state and Kouvel-Fisher method), which validate the correctness of the analysis methods.

\begin{figure}
\includegraphics[width=3.7in]{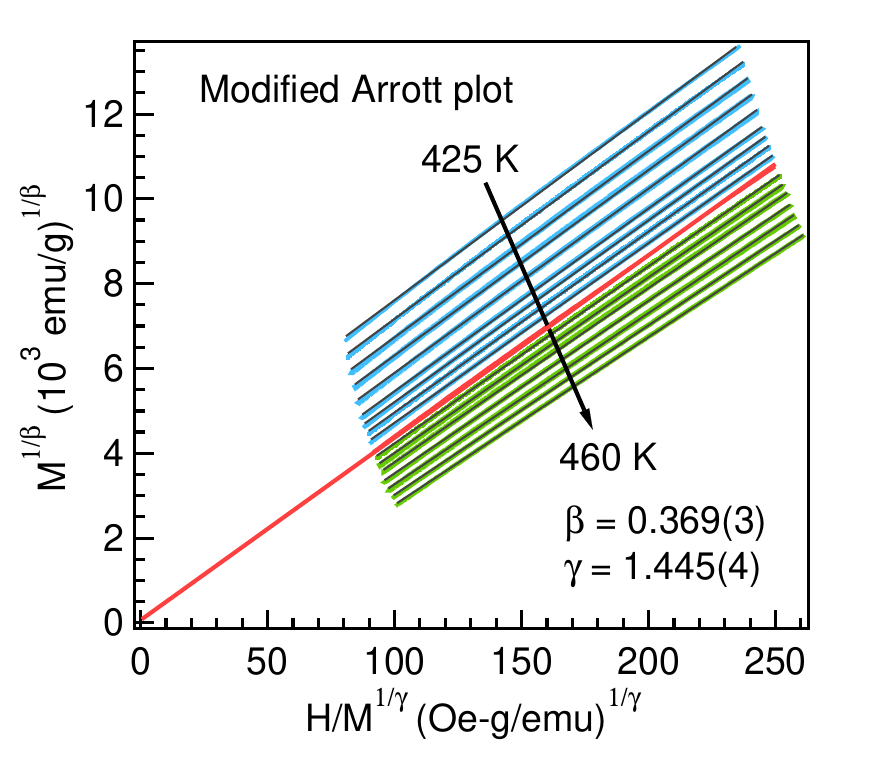}
\caption {The modified Arrott plot (MAP) with $\beta$ = 0.369(3) and $\gamma$ = 1.445(4) values obtained from the K--F method, shown at few temperatures for clarity. The linear fittings are shown with solid black lines in the high field region and the red line at $T_{\rm C}$ is extrapolated up to zero field.}
\label{Fig-Fig_7_Map}
\end{figure}
\begin{figure}[h]
\includegraphics[width=3.6in]{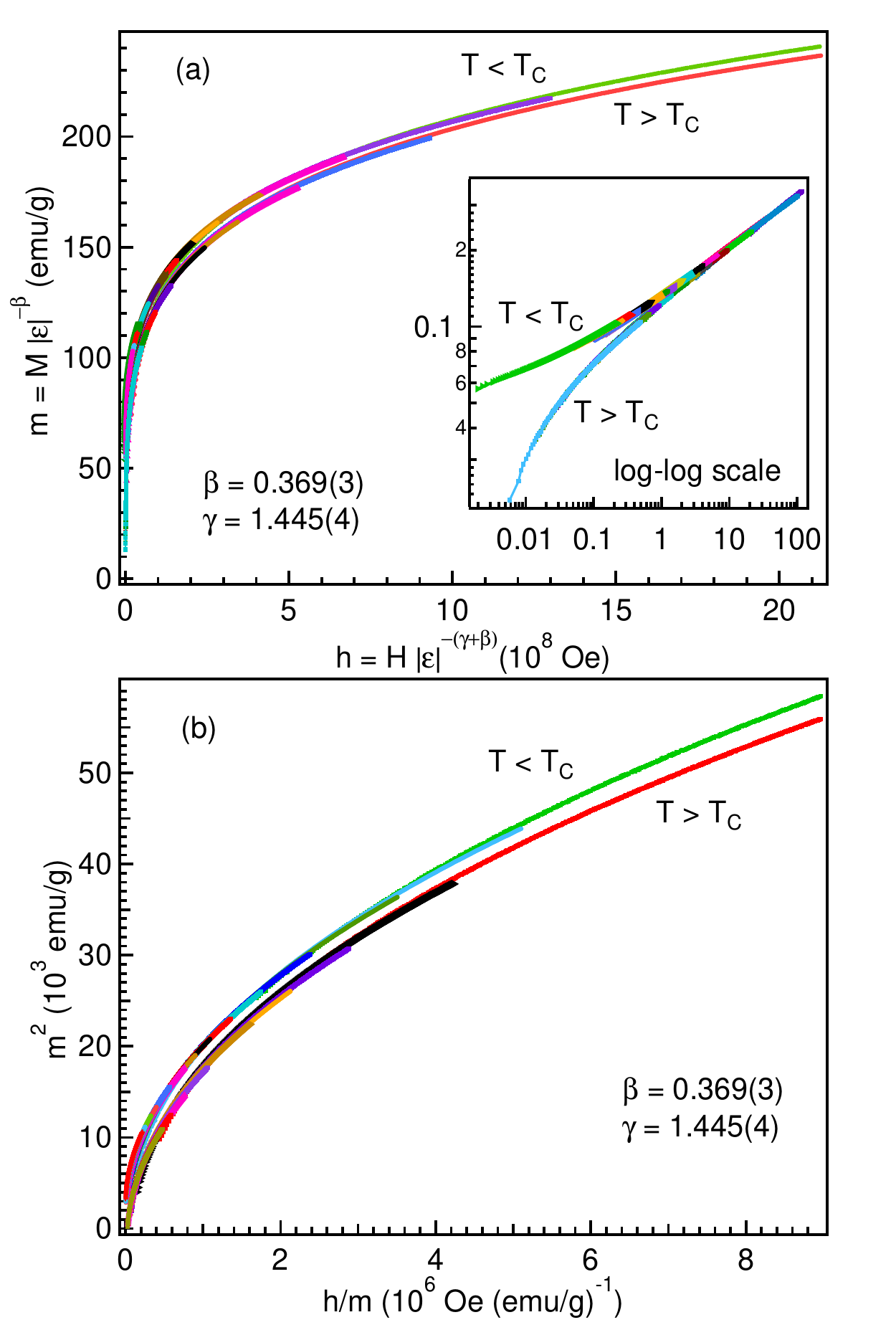}
\caption {(a) The renormalized magnetization (m) as a function of the renormalized field (h), and inset is the log-log plot of m(h) data. (b) m$^2$ versus h/m plot below and above $T_{\rm C}$, using the $\beta$ and $\gamma$ values from the K--F method. The data at different temperatures are represented by various symbols/colors.}
\label{Fig-Fig_8_scaling}
\end{figure}

Moreover, Fig.~\ref{Fig-Fig_6_Ms_beta_gamma}(c) displays the $M-H$ curve corresponding to the $T_{\rm C}$ as well as the inset shows the same in log-log scale. This log$(M)$ against log$(H)$ curve should be linear in the high field region and follow equation~\ref{Eq:Eq_5_delta_TC} where the linear fit (solid red line) in the high-field region gives the slope (1/$\delta$). In addition, the reliability of the critical exponent $\delta$ at the critical isotherm ($M-H$ at 446~K) can be tested by applying Widom scaling formula~\cite{Widom_jcp_65} as written below:
\begin{equation}
\delta = 1 + \frac{\gamma}{\beta} 
\label{Eq:Eq_8_delta}
\end{equation}
Using equation \ref{Eq:Eq_8_delta}, the $\delta$ value can be determined by considering the $\beta$ and $\gamma$ values estimated from Figs.~\ref{Fig-Fig_6_Ms_beta_gamma}(a) and \ref{Fig-Fig_6_Ms_beta_gamma}(b), which yields $\delta$ = 4.956 and 4.961, respectively. These values are quite similar to those found in the critical isotherms, see in Fig.~\ref{Fig-Fig_6_Ms_beta_gamma}(c), within the experimental error. Overall, we find that the extracted values of critical exponents ($\beta$, $\gamma$, $\delta$) as well as the $T_{\rm C}$ are consistent and in good agreement with the Widom scaling relation. The critical exponents obtaining from the K--F method are used to rebuild the MAPs, which manifests straight and parallel lines $\geq$1~T field region, as shown in Fig.~\ref{Fig-Fig_7_Map}. However, the non-linearity at very low field region (not shown) indicates the formation of multi-domain spin structure with different orientations \cite{Nehla_prb_19, Halder_prb_10}.  

Finally, in order to relate the $M(H,\epsilon)$, $H$, and $T_{\rm C}$, the magnetic equation of state can be expressed in the vicinity of $T_{\rm C}$, according to the scaling theory \cite{Pramanik_prb_09, Yan_ssc_18}:
\begin{equation}
M(H, \epsilon)=|\epsilon|^\beta f_{ \pm}\left(\frac{H}{|\epsilon|^{\beta+\gamma}}\right)
\label{Eq:Eq_9_scaling}
\end{equation} 
where the regular functions for temperature, denoted as $f_-$ (for $T<T_{\rm C}$) and $f_+$ (for $T>T_{\rm C}$) play an important role in the scaling equation. In terms of the renormalized magnetization, denoted as $m \approx M(H,\epsilon) |\epsilon|^{-\beta}$, and the renormalized field, denoted as $h \approx H |\epsilon|^{-(\gamma+\beta)}$, the above scaling equation can be modified as $m = f_{\pm}(h)$. The isotherms below $T_{\rm C}$ correspond to the function f$_-$, while for isotherms above $T_{\rm C}$ correspond to the function f$_+$. The plot between $m$ and $h$ align with the two distinct branches of the scaling function when the appropriate values of $\beta$ and $\gamma$ are chosen as described by the equations 8 and 9. This is a crucial test to check whether the critical exponents are stable across the $T_{\rm C}$. In Table~\ref{tab: critical exponents}, we present the values of critical exponents ($\beta$, $\gamma$ and $\delta$) estimated from various approaches, as well as the theoretically anticipated values for various models. Now we can determine whether these exponents satisfy the scaling equation of state (equation~\ref{Eq:Eq_9_scaling}). Therefore, we use the critical exponents from the K--F plot and plotted $m$ versus $h$ in Fig.~\ref{Fig-Fig_8_scaling}(a) and the same plot is zoomed in the low field region in log-log scale, as shown in the inset. It is noted that a few data points are removed from the low field region for the sake of clarity of two distinct branches above and below $T_{\rm C}$. These isotherms clearly exhibit a collapse behavior forming two distinct branches: one located just above and below the $T_{\rm C}$. The accuracy and reliability of the exponents and $T_{\rm C}$ values are further confirmed by a plot between $m^2$ and $h/m$, where $M-H$ curves separate into two branches and follow the scaling function, as shown in Fig.~\ref{Fig-Fig_8_scaling}(b). Here, the crucial point is to investigate the type of magnetic interactions in the proximity of $T_{\rm C}$ to understand the spin interactions. The exchange interaction of long-range extended type itinerant spin interaction spatially decays as $J(r)$$\sim$r$^{-(d+\sigma)}$, where $d$ and $\sigma$ stand for the lattice dimensionality and range of interaction, respectively \cite{Fischer_prb_02, Babu_jpcm_97, Kaul_jmmm_85, Roy_prb_19, Nehla_jap_19}. It has been well established that the value of $\sigma$  is crucial to understand the type of spin interactions; for example $\sigma$\textless 1.5 for the long-range ordering and $\sigma$$\ge$ 2 for the short-range ordering \cite{Fischer_prb_02, Ma_AM23}. Here, in order to compute the value of $\sigma$, the renormalization group technique can be used where the expression for $\gamma$ is given as below \cite{Fisher_prl_72, Fisher_rmp_74}:
		\begin{equation}
			\begin{aligned}
				\gamma= & 1+\frac{4}{d} \left(\frac{n+2}{n+8}\right) \Delta \sigma+\left[\frac{8(n+2)(n-4)}{d^2(n+8)^2}\right] \\
				& \times\left[1+\frac{2 G\left(\frac{d}{2}\right)(7 n+20)}{(n-4)(n+8)}\right] \Delta \sigma^2,
			\end{aligned}
			\label{Eq:Eq_10_spin_interaction}
		\end{equation}
Here the G($d$/2) = 3--$\frac{1}{4}$($\frac{d}{2}$)$^2$, and $\Delta$$\sigma$ = $\sigma$--($n$/2) where $n=$ 3 represents the spin dimensionality. The estimated value of $\gamma$ = 1.445 (see Fig.~7) can be used to determine the value of $\sigma$ using the above equation 12, which found to be $\sim$1.987. Further, we use the value of $\sigma$ to compute the other exponents using the following formulas: $\alpha$ = 2--$\nu$$d$ where the exponent of correlation length $\nu$ = $\gamma$/$\sigma$, $\beta$ = (2--$\alpha$--$\gamma$)/2, and $\delta$ = 1 + ($\gamma$/$\beta$) \cite{Stanley_ox_71, Kaul_jmmm_85}. Here, the determined values of $\beta$ = 0.368 and $\delta$ = 4.927 are found to be consistent with those obtained in the above analysis using the K--F method. In the case of isotropic three-dimensional ferromagnetism, the exchange interaction in the 3D Heisenberg model (with $\beta = 0.365$, $\gamma = 1.386$, and $\delta = 4.8$) exhibits a decay rate that is faster than the power law of $J(r) \sim r^{-5}$ for the value of $\sigma=$ 2. In the present case, the value of $\sigma$ is comparable to 2 and $\beta = 0.369$, which suggest for the the presence of short-range spin interactions and approach to the 3D Heisenberg model  \cite{Roy_prb_19, Ma_AM23}. 

\section{\noindent ~CONCLUSIONS}
		
In summary,  we find that the CoFeV$_{0.8}$Mn$_{0.2}$Si quaternary sample is stable and crystallizes in a single-phase cubic Y-type structure, having $\approx$28\% anti-site disorder identified between V-Si atoms. We observe a second-order paramagnetic to ferromagnetic thermodynamic phase transition at around 446~K. Interestingly, the saturation magnetization of $\approx$2.2~$\mu$$_{\rm B}$/f.u. measured at 5~K is found to be in good agreement with the Slater-Pauling rule, which is perquisite for half-metallic nature. The itinerant ferromagnetic character of moment is obtained from the self-consistent theory and the Rhodes-Wohlfarth ratio of the sample. The values of critical exponents ($\beta$, $\gamma$, and $\delta$) are extracted from the modified Arrott plot, Kouvel--Fisher method, and critical isotherms, which consistently and accurately describe the scaling behavior and suggest for the presence of 3D Heisenberg type interactions in the sample. Furthermore, the spin interaction decays as $J(r)$$\sim$r$^{(-4.99)}$, which indicates the short-range itinerant magnetic ordering. 
		
\section{\noindent ~ACKNOWLEDGMENTS}
		
 GDG thanks the MHRD, India, for the fellowship and IIT Delhi for different experimental facilities; XRD and PPMS in the Department of Physics; glass blowing section, and EDX at the central research facility (CRF). The work is financially supported by the BRNS through DAE Young Scientist Research Award to RSD with the project sanction No. 34/20/12/2015/BRNS. RSD also acknowledges SERB--DST for the financial support through a core research grant (project reference no. CRG/2020/003436). 
		


\end{document}